\documentclass[12pt,preprint]{aastex}

\usepackage{amsmath,amssymb,graphicx,hyperref,cite,twoopt,lineno,color,soul}

\begin{document}

\title{Rapid TeV Gamma-Ray Flaring of BL Lacertae}

\author{
T.~Arlen\altaffilmark{1},
T.~Aune\altaffilmark{2},
M.~Beilicke\altaffilmark{3},
W.~Benbow\altaffilmark{4},
A.~Bouvier\altaffilmark{2},
J.~H.~Buckley\altaffilmark{3},
V.~Bugaev\altaffilmark{3},
A.~Cesarini\altaffilmark{5},
L.~Ciupik\altaffilmark{6},
M.~P.~Connolly\altaffilmark{5},
W.~Cui\altaffilmark{7},
R.~Dickherber\altaffilmark{3},
J.~Dumm\altaffilmark{8},
M.~Errando\altaffilmark{9},
A.~Falcone\altaffilmark{10},
S.~Federici\altaffilmark{11,12},
Q.~Feng\altaffilmark{7},
J.~P.~Finley\altaffilmark{7},
G.~Finnegan\altaffilmark{13},
L.~Fortson\altaffilmark{8},
A.~Furniss\altaffilmark{2},
N.~Galante\altaffilmark{4},
D.~Gall\altaffilmark{14},
S.~Griffin\altaffilmark{15},
J.~Grube\altaffilmark{6},
G.~Gyuk\altaffilmark{6},
D.~Hanna\altaffilmark{15},
J.~Holder\altaffilmark{16},
T.~B.~Humensky\altaffilmark{17},
P.~Kaaret\altaffilmark{14},
N.~Karlsson\altaffilmark{8},
M.~Kertzman\altaffilmark{18},
Y.~Khassen\altaffilmark{19},
D.~Kieda\altaffilmark{13},
H.~Krawczynski\altaffilmark{3},
F.~Krennrich\altaffilmark{20},
G.~Maier\altaffilmark{11},
P.~Moriarty\altaffilmark{21},
R.~Mukherjee\altaffilmark{9},
T.~Nelson\altaffilmark{8},
A.~O'Faol\'{a}in de Bhr\'{o}ithe\altaffilmark{19},
R.~A.~Ong\altaffilmark{1},
M.~Orr\altaffilmark{20},
N.~Park\altaffilmark{22},
J.~S.~Perkins\altaffilmark{23,24},
A.~Pichel\altaffilmark{25},
M.~Pohl\altaffilmark{12,11},
H.~Prokoph\altaffilmark{11},
J.~Quinn\altaffilmark{19},
K.~Ragan\altaffilmark{15},
L.~C.~Reyes\altaffilmark{26},
P.~T.~Reynolds\altaffilmark{27},
E.~Roache\altaffilmark{4},
D.~B.~Saxon\altaffilmark{16},
M.~Schroedter\altaffilmark{4},
G.~H.~Sembroski\altaffilmark{7},
D.~Staszak\altaffilmark{15},
I.~Telezhinsky\altaffilmark{12,11},
G.~Te\v{s}i\'{c}\altaffilmark{15},
M.~Theiling\altaffilmark{7},
K.~Tsurusaki\altaffilmark{14},
A.~Varlotta\altaffilmark{7},
S.~Vincent\altaffilmark{11},
S.~P.~Wakely\altaffilmark{22},
T.~C.~Weekes\altaffilmark{4},
A.~Weinstein\altaffilmark{20},
R.~Welsing\altaffilmark{11},
D.~A.~Williams\altaffilmark{2},
B.~Zitzer\altaffilmark{28} \\
(The VERITAS Collaboration)
}

\author{S.~G.~Jorstad~\altaffilmark{29,30}, N.~R.~MacDonald~\altaffilmark{29}, A.~P.~Marscher~\altaffilmark{29}, P.~S.~Smith~\altaffilmark{31}, R.~C.~Walker~\altaffilmark{32}, T.~Hovatta~\altaffilmark{33}, J.~Richards~\altaffilmark{7}, W.~Max-Moerbeck~\altaffilmark{33}, A.~Readhead~\altaffilmark{33}, M.~L.~Lister~\altaffilmark{7}, Y.~Y.~Kovalev~\altaffilmark{34,35}, A.~B.~Pushkarev~\altaffilmark{36,37}, M.~A.~Gurwell~\altaffilmark{38}, A.~L\"ahteenm\"aki~\altaffilmark{39}, E.~Nieppola~\altaffilmark{39}, M.~Tornikoski~\altaffilmark{39}, E.~J\"arvel\"a~\altaffilmark{39} }

\altaffiltext{1}{Department of Physics and Astronomy, University of California, Los Angeles, CA 90095, USA}
\altaffiltext{2}{Santa Cruz Institute for Particle Physics and Department of Physics, University of California, Santa Cruz, CA 95064, USA}
\altaffiltext{3}{Department of Physics, Washington University, St. Louis, MO 63130, USA}
\altaffiltext{4}{Fred Lawrence Whipple Observatory, Harvard-Smithsonian Center for Astrophysics, Amado, AZ 85645, USA}
\altaffiltext{5}{School of Physics, National University of Ireland Galway, University Road, Galway, Ireland}
\altaffiltext{6}{Astronomy Department, Adler Planetarium and Astronomy Museum, Chicago, IL 60605, USA}
\altaffiltext{7}{Department of Physics, Purdue University, West Lafayette, IN 47907, USA; \href{mailto:qfeng@purdue.edu}{qfeng@purdue.edu}; \href{mailto:cui@purdue.edu}{cui@purdue.edu} }
\altaffiltext{8}{School of Physics and Astronomy, University of Minnesota, Minneapolis, MN 55455, USA}
\altaffiltext{9}{Department of Physics and Astronomy, Barnard College, Columbia University, NY 10027, USA}
\altaffiltext{10}{Department of Astronomy and Astrophysics, 525 Davey Lab, Pennsylvania State University, University Park, PA 16802, USA}
\altaffiltext{11}{DESY, Platanenallee 6, 15738 Zeuthen, Germany}
\altaffiltext{12}{Institute of Physics and Astronomy, University of Potsdam, 14476 Potsdam-Golm, Germany}
\altaffiltext{13}{Department of Physics and Astronomy, University of Utah, Salt Lake City, UT 84112, USA}
\altaffiltext{14}{Department of Physics and Astronomy, University of Iowa, Van Allen Hall, Iowa City, IA 52242, USA}
\altaffiltext{15}{Physics Department, McGill University, Montreal, QC H3A 2T8, Canada}
\altaffiltext{16}{Department of Physics and Astronomy and the Bartol Research Institute, University of Delaware, Newark, DE 19716, USA}
\altaffiltext{17}{Physics Department, Columbia University, New York, NY 10027, USA}
\altaffiltext{18}{Department of Physics and Astronomy, DePauw University, Greencastle, IN 46135-0037, USA}
\altaffiltext{19}{School of Physics, University College Dublin, Belfield, Dublin 4, Ireland}
\altaffiltext{20}{Department of Physics and Astronomy, Iowa State University, Ames, IA 50011, USA}
\altaffiltext{21}{Department of Life and Physical Sciences, Galway-Mayo Institute of Technology, Dublin Road, Galway, Ireland}
\altaffiltext{22}{Enrico Fermi Institute, University of Chicago, Chicago, IL 60637, USA}
\altaffiltext{23}{CRESST and Astroparticle Physics Laboratory NASA/GSFC, Greenbelt, MD 20771, USA.}
\altaffiltext{24}{University of Maryland, Baltimore County, 1000 Hilltop Circle, Baltimore, MD 21250, USA.}
\altaffiltext{25}{Instituto de Astronomia y Fisica del Espacio, Casilla de Correo 67 - Sucursal 28, (C1428ZAA) Ciudad Aut—noma de Buenos Aires, Argentina}
\altaffiltext{26}{Physics Department, California Polytechnic State University, San Luis Obispo, CA 94307, USA}
\altaffiltext{27}{Department of Applied Physics and Instrumentation, Cork Institute of Technology, Bishopstown, Cork, Ireland}
\altaffiltext{28}{Argonne National Laboratory, 9700 S. Cass Avenue, Argonne, IL 60439, USA}

\altaffiltext{29}{Institute for Astrophysical Research, Boston University, USA}
\altaffiltext{30}{St. Petersburg State University, St. Petersburg, Russia}
\altaffiltext{31}{Steward Observatory, University of Arizona, Tucson, AZ 85716, USA}
\altaffiltext{32}{National Radio Astronomy Observatory, PO Box O, Socorro, NM 87801, USA}
\altaffiltext{33}{Cahill Center for Astronomy \& Astrophysics, California Institute of Technology, 1200 E. California Blvd, Pasadena, CA 91125, USA}
\altaffiltext{34}{Astro Space Center of Lebedev Physical Institute, Russian Academy of Sciences, Profsoyuznaya 84/32, 117997 Moscow, Russia}
\altaffiltext{35}{Max-Planck-Institute for Radio Astronomy Auf dem Huegel 69, 53121 Bonn, Germany}
\altaffiltext{36}{Pulkovo Astronomical Observatory, Pulkovskoe Chaussee 65/1, 196140 St. Petersburg, Russia} 
\altaffiltext{37}{Crimean Astrophysical Observatory, 98409 Nauchny, Ukraine}
\altaffiltext{38}{Harvard-Smithsonian Center for Astrophysics, 60 Garden Street, Cambridge, MA 02138, USA}
\altaffiltext{39}{Aalto University Mets\"ahovi Radio Observatory, Mets\"ahovintie 114, FIN-02540 Kylm\"al\"a, Finland}

\begin{abstract}

We report on the detection of a very rapid TeV gamma-ray flare from BL Lacertae on 2011 June 28 with the Very Energetic Radiation Imaging Telescope Array System (VERITAS). The flaring activity was observed during a 34.6-minute exposure, when the integral flux above 200~GeV reached $(3.4\pm0.6) \times 10^{-6} \;\text{photons}\;\text{m}^{-2}\text{s}^{-1}$, roughly 125\% of the Crab Nebula flux measured by VERITAS. The light curve indicates that the observations missed the rising phase of the flare but covered a significant portion of the decaying phase. The exponential decay time was determined to be $13\pm4$ minutes, making it one of the most rapid gamma-ray flares seen from a TeV blazar. The gamma-ray spectrum of BL Lacertae during the flare was soft, with a photon index of $3.6\pm 0.4$, which is in agreement with the measurement made previously by MAGIC in a lower flaring state. Contemporaneous radio observations of the source with the Very Long Baseline Array (VLBA) revealed the emergence of a new, superluminal component from the core around the time of the TeV gamma-ray flare, accompanied by changes in the optical polarization angle. Changes in flux also appear to have occurred at optical, UV, and GeV gamma-ray wavelengths at the time of the flare, although they are difficult to quantify precisely due to sparse coverage. A strong flare was seen at radio wavelengths roughly four months later, which might be related to the gamma-ray flaring activities. We discuss the implications of these multiwavelength results.

\end{abstract}


\section{Introduction}

Blazars form a subclass of active galactic nuclei (AGN) that feature a relativistic jet pointing roughly at the observer. They are known for being highly variable at all wavelengths. In the most extreme cases, the timescales of gamma-ray variability can be as short as a few minutes at very high energies ($\gtrsim 100$ GeV; VHE). Such variability has been detected in several BL Lacertae objects (BL Lacs), including Mrk 421 \citep{1996Natur.383..319G}, Mrk 501 \citep{2007ApJ...669..862A}, and PKS 2155-304 \citep{2007ApJ...664L..71A}, and more recently in the flat-spectrum radio quasar (FSRQ) PKS 1222+21 \citep{2011ApJ...730L...8A}. The rapid variability poses serious challenges to the theoretical understanding of gamma-ray production in blazars. On the one hand, rapid gamma-ray variability implies very compact emitting regions that can be most naturally associated with the immediate vicinity of the central supermassive black hole. On the other hand, the regions must be sufficiently outside the broad-line regions (BLRs) that gamma rays can escape attenuation due to external radiation fields (which, for FSRQs, are particularly strong). Many models have been proposed to resolve these issues \citep{2008MNRAS.386L..28G,2009MNRAS.395L..29G,2011A&A...534A..86T,2012ApJ...749..119B,2012arXiv1202.2123N,2012MNRAS.420..604N}.

The spectral energy distributions (SED) of blazars show two characteristic peaks, with one in the infrared (IR) -- X-ray frequency range and the other in the MeV -- TeV gamma-ray range, respectively. The lower-energy peak is believed to be associated with synchrotron radiation from relativistic electrons in the jet, and the higher-energy peak with inverse-Compton radiation from the same electrons in leptonic models; the situation is more complex in hadronic models. Going from high-power quasars to low-power BL Lacs, the peaks shift systematically to higher frequencies. Most of the known TeV gamma-ray blazars are BL Lacs. They have been historically divided into high-frequency-peaked BL Lacs (HBLs) and low-frequency-peaked BL Lacs (LBLs) \citep{1995ApJ...444..567P,1998MNRAS.299..433F}. BL Lacertae, the archetypical source of the class, is an LBL in this classification scheme.

BL Lacertae (also known as 1ES 2200+420) is an AGN located at a redshift of $z=0.069$ \citep{1978ApJ...219L..85M}. In 1998, the Crimean Observatory reported a detection of the source at $>$100\% of the Crab Nebula flux above 1 TeV \citep{2001ARep...45..249N}. Subsequently, the MAGIC Collaboration reported another detection during an active state in 2005, but at a much lower flux level (only about 3\% of the Crab Nebula flux) \citep{2007ApJ...666L..17A}. Triggered by activities seen with the {\it Fermi} LAT \citep{Atel3368} and AGILE \citep{Atel3387} at GeV gamma-ray energies, as well as in the optical \citep{Atel3371}, near-IR \citep{Atel3375}, and radio \citep{Atel3380} in 2011 May, we began to monitor BL Lacertae more regularly at TeV gamma-ray energies with VERITAS. In this work, we report the detection of a rapid, intense VHE gamma-ray flare from the direction of the source on MJD 55740 (2011 June 28), as well as the results from the multiwavelength observations that were conducted around the time of this flare. 

\section{Observations and Data Analysis}
\subsection{Very High Energy Gamma Ray}
VERITAS is an array of four 12-meter imaging atmospheric Cherenkov telescopes located in southern Arizona. Each telescope is equipped with a focal-plane camera with 499 photomultiplier tubes, covering a 3.5$^{\circ}$ field-of-view \citep{2008AIPC.1085..657H}. VERITAS is sensitive to VHE radiation in the energy range from $\sim$100~GeV to $\sim$30~TeV, being capable of making a detection at a statistical significance of 5 standard deviations (5 $\sigma$) of a point source of 1\% of the Crab Nebula flux in $\sim$25~hours.

Prior to the intensified monitoring campaign with VERITAS, BL Lacertae had also been observed on a number of occasions, mostly with the full array. The data from those observations are also used in this work to establish a longer baseline. The total live exposure time (after quality selection) amounts to 20.3 hrs from 2010 September to 2011 November, with zenith angles ranging from 10 to 40 degrees. The source was not detected throughout the time period, except for one night on MJD 55740 (2011 June 28), when the automated realtime analysis revealed the presence of a rapidly flaring gamma-ray source in the direction of BL Lacertae. On that night, BL Lacertae was observed only with three telescopes in the ``wobble'' mode \citep{2001A&A...370..112A} with 0.5 degrees offset, because one telescope was temporarily out of commission. Starting at 10:22:24 UTC, two 20-minute runs were taken on the source under good weather conditions, with the zenith angle varying between 10 and 13~degrees. No additional runs were possible due to imminent sunrise. The total live exposure time was 34.6~minutes. 

The data were analyzed using the data analysis package described in \citet{2008ICRC....3.1385C}. The analysis procedure includes raw data calibration, image parameterization \citep{1985ICRC....3..445H}, event reconstruction, background rejection and signal extraction \citep{2008ICRC....3.1325D}. The standard data quality cuts (identical for the four- and three-telescope configuration), which were previously optimized for a simulated soft point source of $\sim$6.6\% of the Crab Nebula flux at 200 GeV and a photon index of 4, were applied to the shower images. The cuts used were: an integrated charge lower cut of 45 photoelectrons, a distance (between the image centroid and the center of the camera) upper cut of 1.43~degrees, a minimum number of pixels cut of 5 for each image, inclusive, mean scaled width and length cuts $0.05<\text{MSW}<1.15$, and $0.05<\text{MSL}<1.3$, respectively. A cut of $\theta^2<0.03 \;\text{deg}^2$ on the size of the point-source search window was made, where $\theta$ is the angle between the reconstructed gamma-ray direction and the direction to the source. A specific effective area corresponding to these cuts and the relevant array configuration was generated from simulations and was used to calculate the flux. The reflected-region background model \citep{2007A&A...466.1219B} was applied for background estimation, a generalized method from \citet{1983ApJ...272..317L} was used for the calculation of statistical significance, and upper limits were calculated using the method described by \citet{2005NIMPA.551..493R}. The results were confirmed by an independent secondary analysis with a different analysis package, as described in \citet{2008ICRC....3.1325D}. 

\subsection {High Energy Gamma Ray}

The {\it Fermi} Large Area Telescope (LAT) is a pair-conversion high-energy gamma-ray telescope covering an energy range from about 20 MeV to more than 300 GeV \citep{2009ApJ...697.1071A}. It has a large field-of-view of 2.4~sr, and an effective area of $\sim8000 \; \text{cm}^2$ for $>1$~GeV. In its nominal (survey) mode, the {\it Fermi}-LAT covers the full sky every 3~hours. 

During the time window when VERITAS detected a rapid flare on MJD 55740 (2011 June 28), BL~Lacertae was in the field of view of the LAT for about 16 minutes (MJD 55740.431 - 55740.442). In analyzing the simultaneous LAT data, we selected {\em Diffuse} class photons with energy between 0.2 and 10\,GeV in a $16^{\circ} \times 16^{\circ}$ region of interest (ROI) centered at the location of BL~Lacertae. Only events with rocking angle $< 52^{\circ}$ and zenith angle $< 100^{\circ}$ were selected. The data were processed using the publicly available {\it Fermi}-LAT tools (\texttt{v9r23p1}) with standard instrument response functions (\texttt{P7SOURCE\_V6}). For such a short exposure, a very simple model containing the source of interest and the contribution of the galactic (using file \texttt{gal\_2yearp7v6\_v0.fits}) and isotropic (using file \texttt{iso\_p7v6source.txt}) diffuse emission was used. The contribution of the other known gamma-ray sources in the ROI is assumed to be negligible compared to that of BL~Lacertae and the diffuse emission.

The model is fitted to the data using a binned likelihood analysis (\texttt{gtlike}), where the only free parameters are the spectral normalization and the power-law index of BL~Lacertae. The contribution of the galactic and isotropic diffuse emission was fixed to a normalization of 1.0, which is compatible with the values obtained when analyzing the same field of view during longer timespans. The results are used to construct an energy spectrum of BL~Lacertae. We also performed an unbinned likelihood analysis and obtained similar spectral results.

For comparison, we repeated the analyses for a longer period (of 24 hours) centered at the time of the VERITAS observations, as well as for times prior to the VERITAS-detected flare (between 2011 May 26 and 2011 June 26, or MJD 55707--55738). For the latter, we adopted a source model that incorporates all sources in the 2FGL catalog within the ROI and within 5~degrees of the ROI edges. The spectral results were extracted by adopting a custom spectral code (\texttt{SED\_scripts}) available on the {\it Fermi}-LAT website. In all cases, the LAT spectrum of BL Lacertae can be well described by a power law, which justifies the assumption made in the likelihood analyses.

A daily-binned light curve integrated above 0.1~GeV was derived covering the period MJD 55652-55949 (2011 April 01 - 2012 Jan 23) using the likelihood method described above. In each 1-day bin, the flux and the corresponding 1$\sigma$ error are calculated if the test statistic (TS) value is greater than 1, otherwise an upper limit is calculated.  

\subsection {X-ray and Ultraviolet}

BL Lacertae was also observed with the XRT and UVOT instruments on board the {\it Swift} satellite \citep{2004ApJ...611.1005G} contemporaneously with the gamma-ray flare in 18 exposures between MJD 55704 (2011 May 23) and MJD 55768 (2011 July 26), including six $\sim$2 ks Target of Opportunity (ToO) observations on six nights following the VHE flare on MJD 55740. The combination of the X-ray telescope (XRT) and UV/optical telescope (UVOT) provided useful coverage in soft X-rays and UV, although none of the observations were simultaneous with the VERITAS observations during the flare.

We analyzed the XRT data using the HEASOFT package (version 6.11). The event files are calibrated and cleaned using the calibration files from 2011 September 5. The data were taken in the photon-counting (PC) mode, and were selected from grades 0 to 12 over the energy range 0.3-10~keV. Since the rates did not exceed 0.5 counts per second, pile-up effects were negligible. Source counts were extracted with a 20 pixel radius circle centered on the source, while background counts were extracted from a 40 pixel radius circle in a source-free region. Ancillary response files were generated using the \texttt{xrtmkarf} task, with corrections applied for the point-spread function (PSF) losses and CCD defects. The corresponding response matrix from the XRT calibration files was applied. The spectrum was fitted with an absorbed power law model, allowing the neutral hydrogen (HI) column density ($N_H$) to vary. The best fitted value of $N_H$ is $(0.24\pm 0.01) \times10^{22} \;\text{cm}^{-2}$, which is in agreement with the result of $N_H = 0.25  \times10^{22} \;\text{cm}^{-2}$ presented by \citet{2003A&A...408..479R}, but is larger than the value of $N_H = 0.18  \times10^{22} \;\text{cm}^{-2}$ from the Leiden/Argentine/Bonn (LAB) survey of galactic HI \citep{2005A&A...440..775K}.

The UVOT cycled through each of the optical and the UV pass bands V, B, U, UVW1, UVM2 and UVW2. Data were taken in the {\it image mode} discarding the photon timing information. Only data from UVW2 band are shown in this work; the other bands roughly track UVW2.
The photometry was computed using an aperture of $5^"$ following the general prescription of \citet{2008MNRAS.383..627P} and \citet{2010MNRAS.406.1687B}. Contamination by background light arising from nearby sources was removed by introducing {\it ad hoc} exclusion regions. 
Adopting the $N_H$ value provided by the XRT analysis and assuming $E(B-V)=0.34$~mag \citep{1997IAUC.6700....3M}, we estimated $R_V=3.2$ \citep{2009MNRAS.400.2050G}. Then, the optical/UV galactic extinction coefficients were applied \citep{1999PASP..111...63F}. The host galaxy contribution has been estimated using the PEGASE-HR code \citep{2004A&A...425..881L} extended for the ultraviolet UVOT filters. Moreover, there is no pixel saturation in the source region and no significant photon loss. Therefore, it is possible to constrain the systematics to below 10\%.

\subsection {Optical}

As part of the Steward Observatory spectropolarimetric monitoring project~\citep{2009arXiv0912.3621S}, BL Lacertae was observed regularly with the $2.3$m Bok Telescope and the $1.54$m Kuiper Telescope in Arizona. Measurements of the V-band flux density and optical linear polarization are from the Steward Observatory public data archive (\url{http://james.as.arizona.edu/~psmith/Fermi/}). The data were reduced and calibrated following the procedures described by \citet{2009arXiv0912.3621S}. We note that there is a 180-degree degeneracy in polarization angle, so we shifted some polarization angles by 180~degrees to minimize the change between two consecutive measurements. No corrections to the data have been made for the contribution from the host galaxy, or interstellar polarization, extinction and reddening. However, these issues have little effect on variability studies.

\subsection {Radio}

BL Lacertae was observed with the VLBA at 43 GHz, roughly once a month, as part of the monitoring program of gamma-ray bright blazars at Boston University (\url{http://www.bu.edu/blazars/VLBAproject.html}). Two extra epochs of imaging were added via Director's Discretionary Time on 2011 July 6 and 29. The data were correlated at the National Radio Astronomy Observatory in Socorro, NM, and then analyzed at Boston University following the procedures outlined by \citet{2005AJ....130.1418J}. The calibrated total and polarized intensity images were used to investigate the jet kinematics and to calculate the polarization parameters (degree of polarization $p$ and position angle of polarization $\chi$) for the whole source imaged at 43~GHz with the VLBA and for individual jet components. The uncertainties of polarization parameters were computed based on the noise level of total and polarized intensity images and do not exceed 0.6\% and 3.5~degrees for degree of polarization and position angle of polarization, respectively.

BL Lacertae is also in the sample of the Monitoring Of Jets in Active galactic nuclei with VLBA Experiments (MOJAVE) program. For this work, we only used results from polarization measurements at 15.4 GHz. The data reduction procedures are described by \citet{2009AJ....137.3718L}. Briefly, the flux density of the core component is derived from a Gaussian model fit to the interferometric visibility data. Polarization properties of the core are then derived by taking the mean Stokes Q and U flux densities of the nine contiguous pixels that are centered at the Gaussian peak pixel position of the core fit. The results include fractional linear polarization, electric vector position angle (note the 180-degree degeneracy), and polarized flux densities. The flux density has an uncertainty of $\sim 5\%$, and the position angle of polarization has an uncertainty of $\sim 3$~degrees.

For better sampling, we used data from blazar monitoring programs with the Owens Valley Radio Observatory (OVRO) at 15.4~GHz, with the Mets\"ahovi Radio Observatory (MRO) at 37~GHz, and with the Submillimeter Array (SMA) at 230 and 350~GHz, respectively.
The OVRO 40~m uses off-axis dual-beam optics and a cryogenic high electron mobility transistor (HEMT) low-noise amplifier with a 15.0~GHz center frequency and 3~GHz bandwidth. The two sky beams are Dicke-switched using the off-source beam as a reference, and the source is alternated between the two beams in an ON-ON fashion to remove atmospheric and ground contamination. Calibration is achieved using a temperature-stable diode noise source to remove receiver gain drifts and the flux density scale is derived from observations of 3C~286 assuming the \citet{1977A&A....61...99B} value of 3.44~Jy at 15.0~GHz. The systematic uncertainty of about 5\% in the flux density scale is not included in the error bars. Complete details of the reduction and calibration procedure are found in \citet{2011ApJS..194...29R}.

The 37 GHz observations were made with the 13.7 m diameter Mets\"ahovi radio telescope, which is a radome-enclosed paraboloid antenna situated in Finland (24 23' 38''E, +60 13' 05''). The measurements were made with a 1 GHz-band dual beam receiver centered at 36.8 GHz. The observations are ON--ON observations, alternating the source and the sky in each feed horn. A typical integration time to obtain one flux density data point is between 1200 s and 1400 s. The detection limit of the telescope at 37 GHz is on the order of 0.2 Jy under optimal conditions. Data points with a signal-to-noise ratio $< 4$ are treated as non-detections.
The flux density scale is set by observations of DR 21. Sources NGC 7027, 3C 274 and 3C 84 are used as secondary calibrators. A detailed description of the data reduction and analysis is given in \citet{1998A&AS..132..305T}. The error estimate in the flux density includes the contribution from the measurement rms and the uncertainty of the absolute calibration.

Observations of BL~Lacertae at frequencies near 230 and 350~GHz are from the Submillimeter Array (SMA), a radio interferometer consisting of eight 6-m diameter radio telescopes located just below the summit of Mauna Kea, Hawaii. These data were obtained and calibrated as part of the normal monitoring program initiated by the SMA \citep[see][]{2007ASPC..375..234G}. Generally, the signal-to-noise ratio of these observations exceeds 50 and is often well over 100, and the true error on the measured flux density is limited by systematic rather than signal-to-noise effects. Visibility amplitudes are calibrated by referencing to standard sources of well-understood brightness, typically solar system objects such as Uranus, Neptune, Titan, Ganymede, or Callisto. Models of the brightness of these objects are accurate to within around 5\% at these frequencies. Moreover, the SMA usually processes only a single polarization at one time, and there is evidence that BL~Lacertae in 2011 exhibited a fairly strong ($\sim$15\%) linear polarization. For a long observation covering a significant range of parallactic angle, the effect of the linear polarization would be largely washed out, providing a good measure of the flux density. However, not all observations of BL Lacertae covered a significant range of parallactic angle, and thus in some cases we would expect a potential absolute systematic error up to $10$\%. In most cases, we expect that the total systematic error is around 7.5\%. 

\section{Results}
\subsection{Gamma Ray Properties}
The VERITAS analysis showed an excess of 212 $\gamma$-like events, corresponding to $11.0 \pm 0.8 \; \gamma/\text{min}$ and a 21.1$\sigma$ detection of BL Lacertae in the first observation run on MJD 55740 (2011 June 28), with an effective exposure of 19.3 minutes starting at 10:22:24 UTC. The second run, with an effective exposure of 15.3 minutes, yielded an excess of only 33 $\gamma$-like events, corresponding to a 4.1$\sigma$ detection. The VERITAS analysis of 19.7-hour data from 2010 September to 2011 November, excluding the two flaring runs, showed an excess of 21 $\gamma$-like events, and a statistical significance of 0.28$\sigma$.

\begin{figure}[here]
\includegraphics[width=0.9\textwidth]{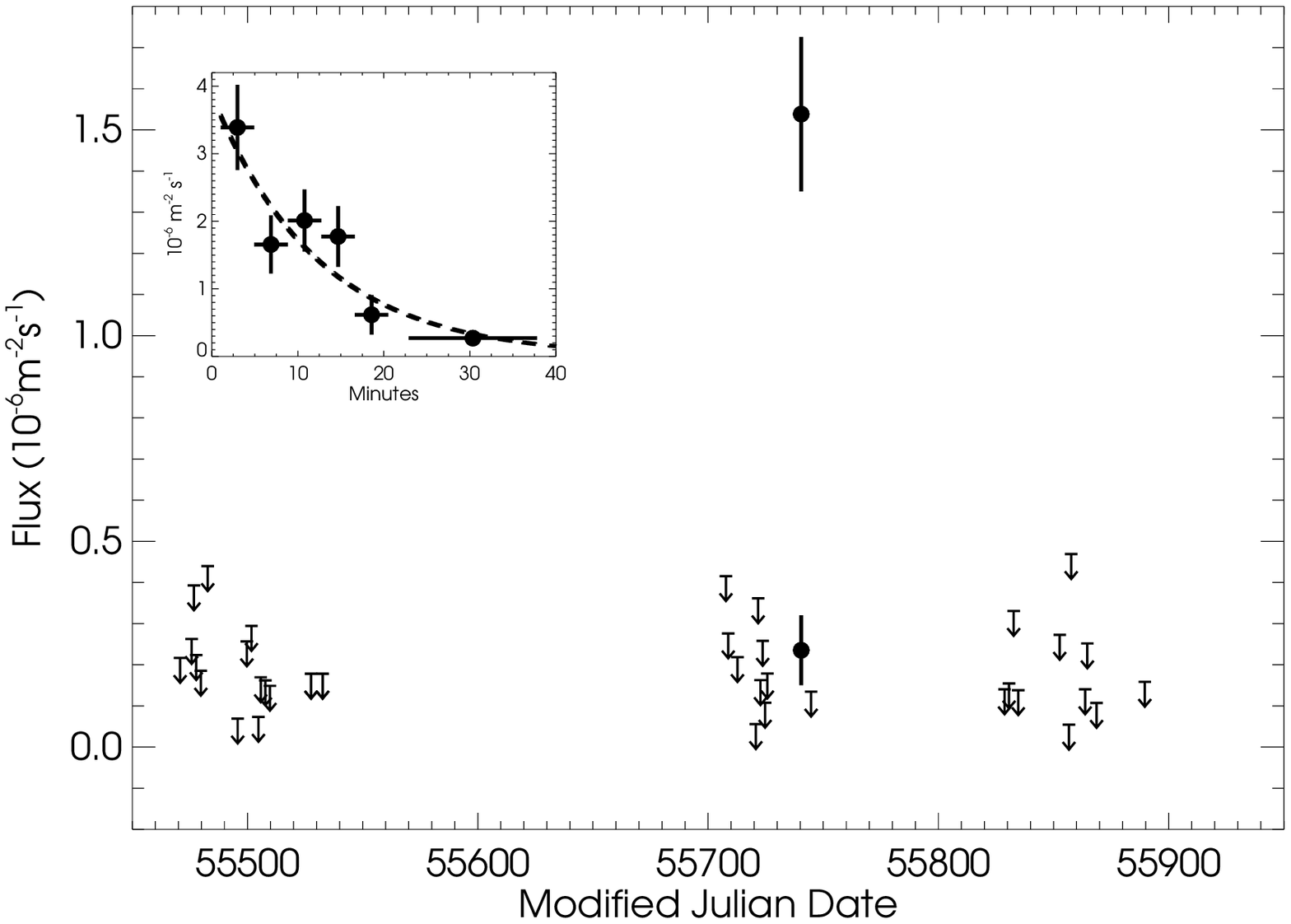}
\caption{TeV gamma-ray light curve of BL Lacertae ($>200$ GeV). When the source was not significantly detected, 99\% confidence upper limits are shown. The upper limits were derived by combining data from all observation runs for each night, but for the night of the flare, the fluxes derived from the two individual runs are shown separately. The inset shows the flare in detail, in 4-minute bins for the first run, and one 16-minute bin for the second run, with minute 0 indicating the start of the first run. The dashed line shows the best fit to the profile with an exponential function (see text). }
\label{Fig TeV}
\end{figure}

Focusing on the two flaring runs, we produced a light curve with 4-minute bins as shown in the inset of Fig.~\ref{Fig TeV}. The fluxes were computed with a lower energy threshold of 200 GeV. The observations missed the rising phase of the flare. In 4-minute bins, the highest flux that was measured is $(3.4\pm0.6) \times 10^{-6} \;\text{photons}\;\text{m}^{-2}\text{s}^{-1}$, which corresponds to about 125\% of the Crab Nebula flux above 200 GeV, as measured with VERITAS. To quantify the decay time, the light curve was fitted with an exponential function $I(t)=I_0\times \exp{(-t/\tau_d)}$, and the best-fit decay time was $\tau_d=13\pm 4 \; \text{minutes}$.

To determine the position of the flaring gamma-ray source, we fitted a 2-d Gaussian function to the uncorrelated map (binned to 0.05$^{\circ}$) of excess events (after acceptance correction) from both runs. The best-fit right ascension and declination (J2000) are $22^h02^m37^s$ and $+42^{\circ} 15^{'} 25^{"}$, respectively, with a statistical uncertainty of $\sim$0.01$^{\circ}$ along both directions. The source is thus named as VER J2202+422. According to the Simbad database, BL Lacertae is the only object within a radius of $2^{'}$.  

\begin{figure}[here]
\includegraphics[width=0.9\textwidth]{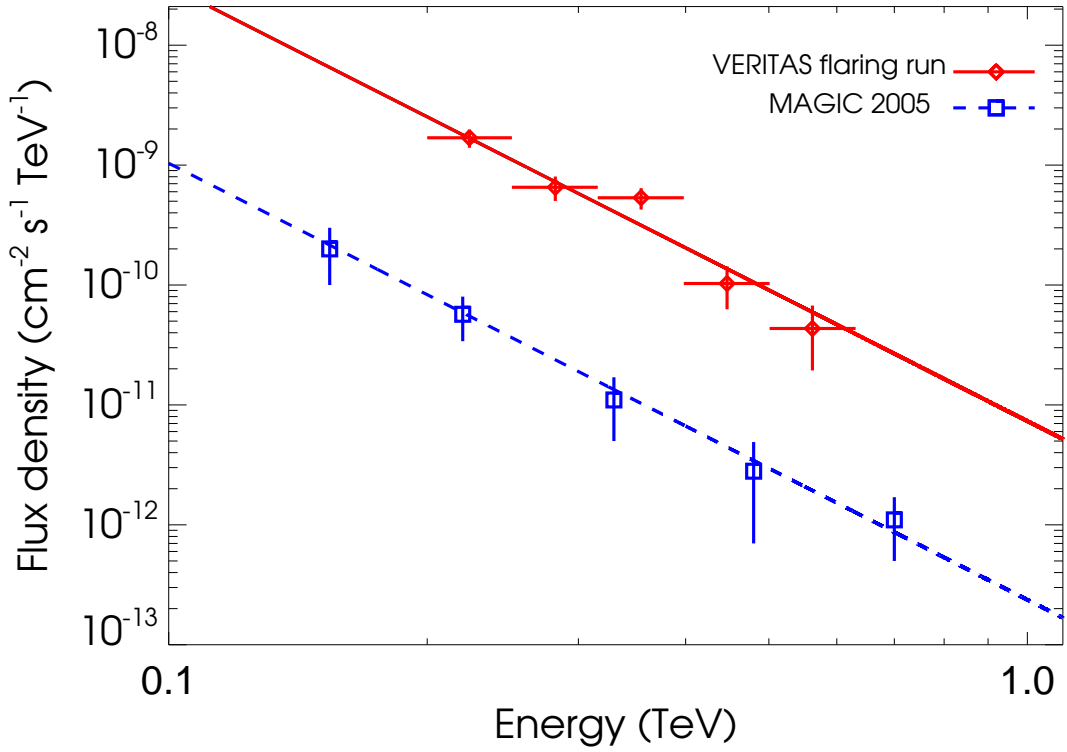}
\caption{TeV gamma-ray spectrum of BL Lacertae. The VERITAS data points are shown as red open diamonds, along with the best-fit power law (solid line). For comparison, we also show the published MAGIC spectrum of the source as blue open squares, along with the best-fit power law (dashed line). The two power laws have comparable slopes.}
\label{Fig OverSpec}
\end{figure}

\begin{figure}[here]
\includegraphics[width=0.9\textwidth]{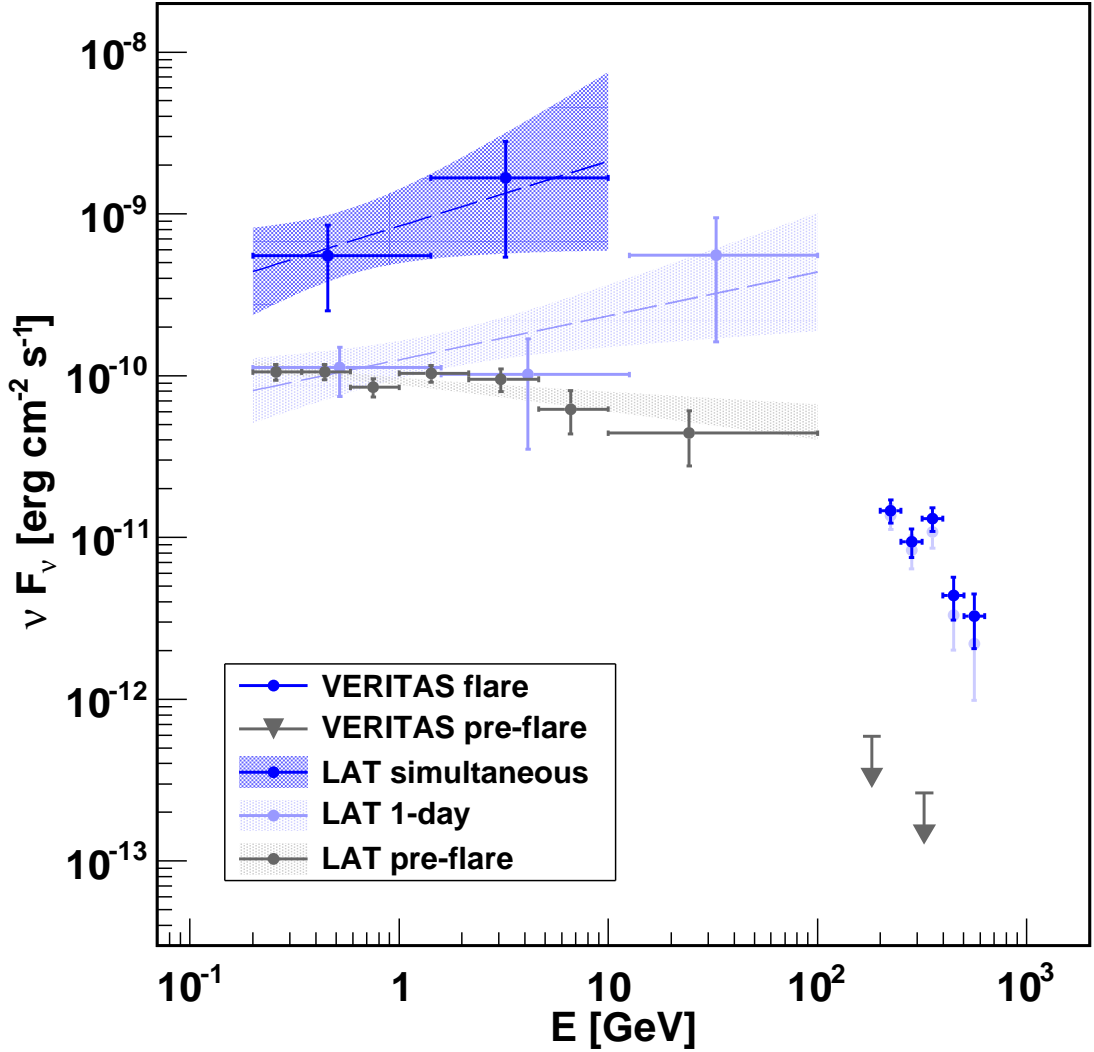}
\caption{Broadband gamma-ray spectrum of BL Lacertae. For VERITAS, the flare spectrum with EBL correction is shown as blue points with uncertainties, and the spectrum without EBL correction is shown as lighter blue points. Also plotted are the 95\% confidence upper limits that were derived from 14 VERITAS observation runs in the month prior to the flare. For the {\it Fermi}-LAT data, the simultaneous, 1-day, and pre-flare spectra are shown in blue, light blue, and gray. The pre-flare spectrum was derived from LAT observations taken over a period of one month before the TeV gamma-ray flare. The shaded areas show the 1$\sigma$ confidence interval of the overall Fermi derived spectra, independent from the data points.}
\label{Fig FermiVER}
\end{figure}

Using data from the first flaring run, we extracted a gamma-ray spectrum (Fig.~\ref{Fig OverSpec}). It can be fitted with a power law,
\begin{equation}
dN/dE=(0.58\pm 0.07)\times 10^{-9} \times (E/0.3\text{TeV})^{(-3.6 \pm 0.4)} \text{cm}^{-2} \text{s}^{-1} \text{TeV}^{-1}.
\end{equation} 
Also shown in the figure is the gamma-ray spectrum of BL Lacertae obtained with MAGIC during a lower-flux state \citep{2007ApJ...666L..17A}. The two gamma-ray spectra have nearly the same slope. This is in contrast with the typical spectral hardening trend of a flaring blazar~\citep[e.g.,][]{1990ApJ...356..432G}. It may reflect the fact that in both cases the TeV gamma rays fall on the steeply falling part of the high-energy SED peak, which might not be sensitive to flux changes. A spectrum was also constructed from both runs and fitted with a power law: 
\[
dN/dE=(0.30\pm 0.03)\times 10^{-9} \times (E/0.3\text{TeV})^{(-3.8 \pm 0.3)} \text{cm}^{-2} \text{s}^{-1} \text{TeV}^{-1}.
\]
To better constrain the gamma-ray SED, we plotted the {\it Fermi}-LAT spectra of the source averaged over several time periods (16 minutes, one day and one month) along with the VERITAS spectra, in Fig.~\ref{Fig FermiVER}. The VERITAS spectra during the flare both with and without extragalactic background light (EBL) corrections \citep{2011MNRAS.410.2556D} are shown, as well as the 95\% confidence upper limits from 14 observations in one month before the flare. The best-fit LAT results for the 16 minutes simultaneous with VERITAS converge to $F_{0.2-10\,\mathrm{GeV}} = \left(2.1 \pm 0.9\right)\, \times 10^{-6}\mathrm{cm}^{-2}\mathrm{s}^{-1}$ with a spectral index of $\Gamma = 1.6 \pm 0.4$, with a test statistic of 35. The 16-minute exposure is very short for the LAT, so the uncertainty is large. Together, the simultaneous VERITAS and LAT spectra show that the gamma-ray SED peak probably lies between 10 and 100~GeV. Moreover, the LAT results provide evidence for spectral hardening during the VERITAS flare, with the best-fit photon index changing from about $2.12\pm 0.05$ to $1.6\pm 0.4$ in the LAT band; note, however, the large uncertainties.

\subsection{Multiwavelength Properties}

\begin{figure}[here]
\includegraphics[width=0.9\textwidth]{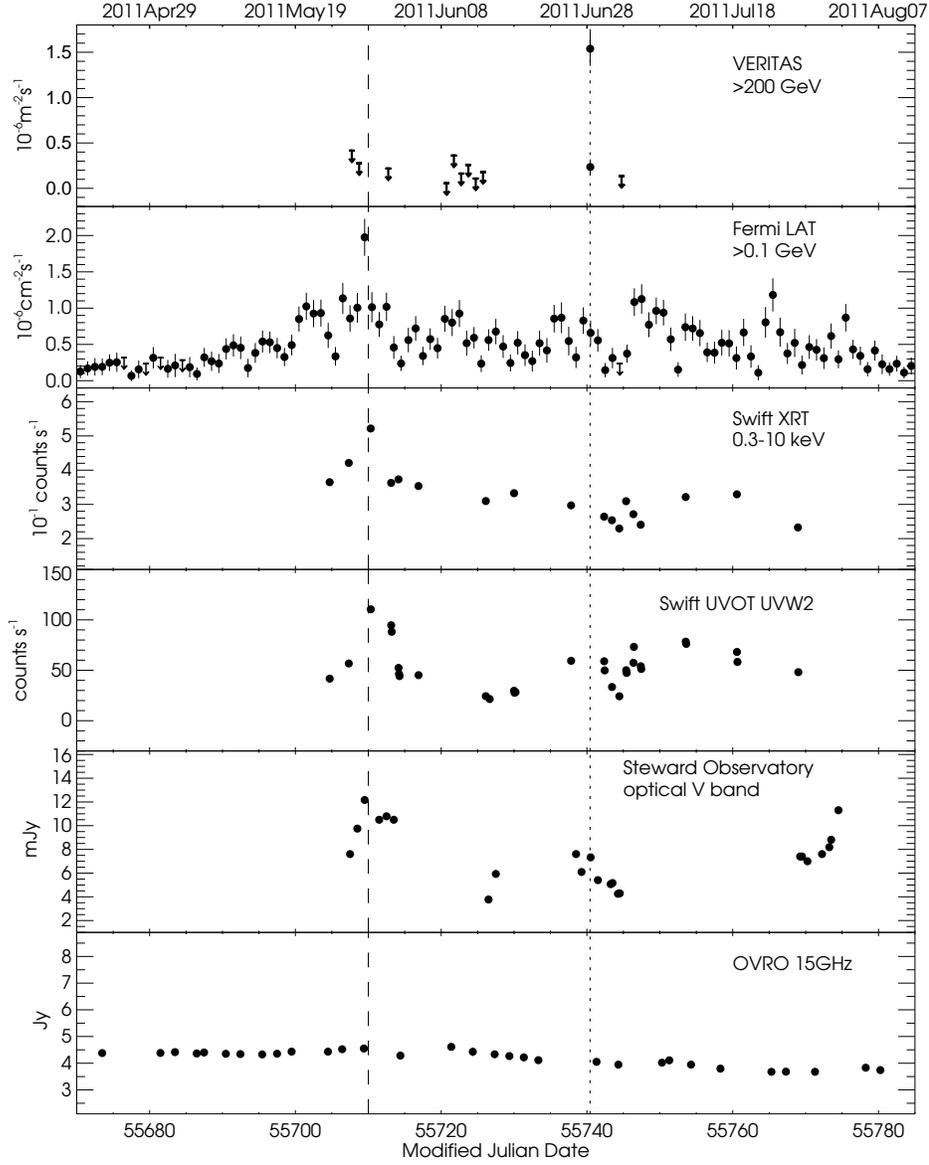}
\caption{Multiwavelength light curves of BL Lacertae. {\it Fermi}-LAT 1$\sigma$ upper limits are shown as arrows. The dotted line indicates the time of the TeV gamma-ray flare seen with VERITAS, while the dashed line shows the time of a rapid GeV gamma-ray flare seen with the {\it Fermi}-LAT.}
\label{Fig flc}
\end{figure}

Fig.~\ref{Fig flc} shows the multiwavelength light curves of BL Lacertae. The TeV gamma-ray flare occurred when the source was active and variable at GeV energies (\citealt{Atel3368}). Although the LAT light curve shows variability on a timescale of days, no rapid flaring on shorter timescales is apparent. On the other hand, the LAT could have missed a flare as rapid as the TeV flare, due to the lack of statistics. There was no apparent activity in the soft X-ray either, but changes in flux may have occurred at UV and optical wavelengths (although there was no UVOT coverage during the TeV gamma-ray flare). The XRT spectrum was relatively hard, with a photon index of $\sim$1.8, compared to that obtained during the {\it Fermi}-LAT campaign during a low state in 2008 \citep{2011ApJ...730..101A}. There was no apparent variation in the radio flux from the source at the time of the TeV gamma-ray flare. It is interesting to note that the LAT caught a very intense rapid flare earlier (around MJD 55710) that was accompanied by similar flaring activities at soft X-ray, UV, and optical wavelengths. Unfortunately, the source was not observed with VERITAS at that time.

\begin{figure}[here]
\includegraphics[width=0.9\textwidth]{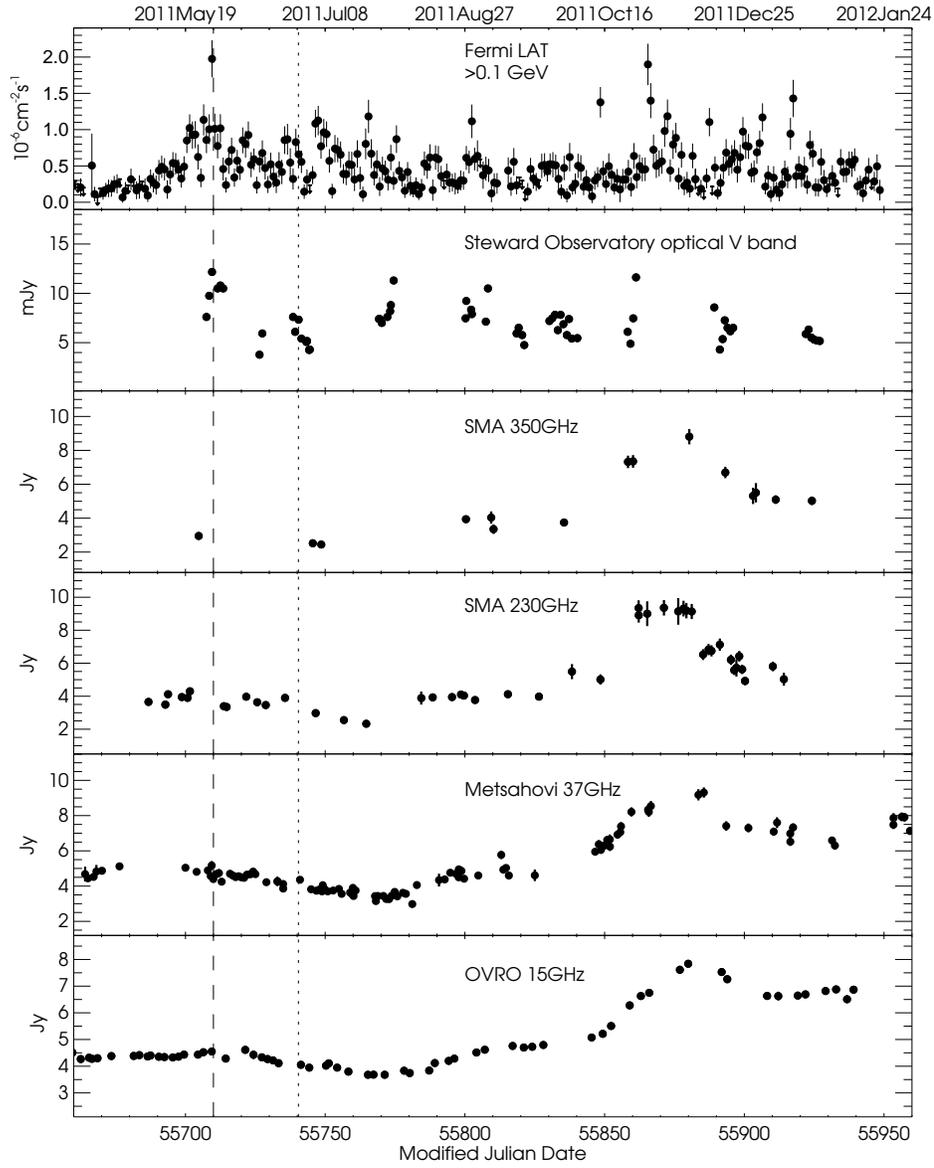}
\caption{Extended {\it Fermi}-LAT, optical, and radio light curves of BL Lacertae. As in Fig.~\ref{Fig flc}, the dotted line indicates the time of the VERITAS flare, and the dashed line shows the time of the {\it Fermi}-LAT flare. }
\label{Fig radiolc}
\end{figure}

\begin{figure}[here]
\includegraphics[width=0.9\textwidth]{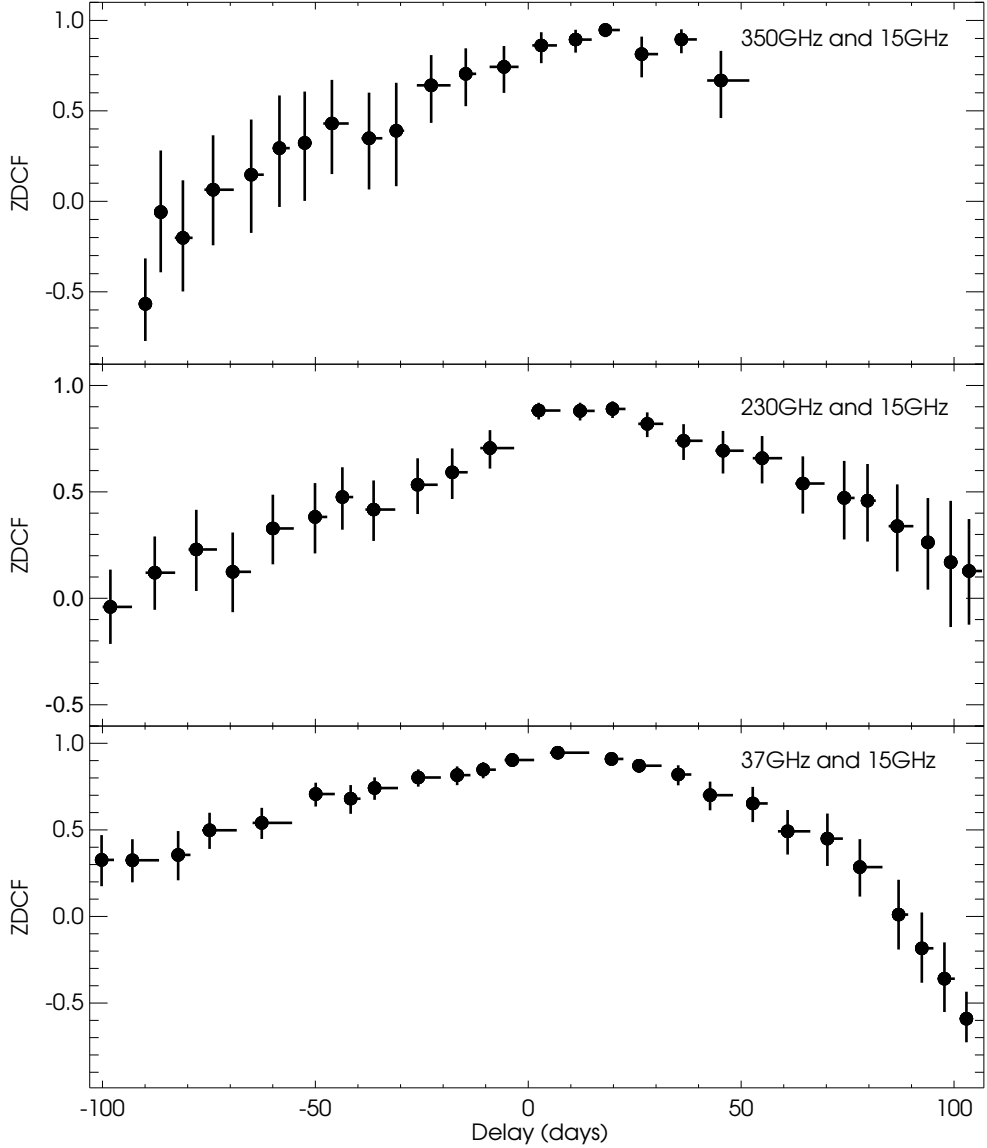}
\caption{Cross-correlation functions derived from the light curves of BL Lacertae: {\it top} between the SMA 350 GHz and OVRO 15 GHz bands; {\it middle} between the SMA 230 GHz and OVRO 15 GHz bands; and {\it bottom} between the Mets\"ahovi 37 GHz and OVRO 15GHz bands. Positive delays indicate ``lead'' with respect to the reference (OVRO 15 GHz) band. }
\label{Fig ZDCF}
\end{figure}

Extending the {\it Fermi}-LAT, optical, and radio light curves to later times, we clearly see an intense flare that occurred at 15.4~GHz, 37~GHz, and 230~GHz, as shown in Fig.~\ref{Fig radiolc}, about four months after the TeV gamma-ray flare. Although the elevated flux is also evident at 350~GHz, the flare is poorly sampled. 
The well-sampled {\it Fermi}-LAT light curve indicates some elevated and variable GeV emission in 2011 November. However, the presence of similar GeV variabilities from 2011 May to the end of the year makes it difficult to establish a correlation between GeV and radio bands. 
We cross-correlated the light curves at the four radio frequencies, using the $z$-transformed discrete correlation function (ZDCF) \citep{1997ASSL..218..163A}. The results are shown in Fig.~\ref{Fig ZDCF}, indicating high degree of correlation among the bands. From the ZDCFs, the corresponding time lags were measured, using a publicly available likelihood code (\texttt{PLIKE}), and are plotted against $\nu^{-1}$ in Fig.~\ref{Fig TFreq}. 

\begin{figure}[here]
\includegraphics[width=0.9\textwidth]{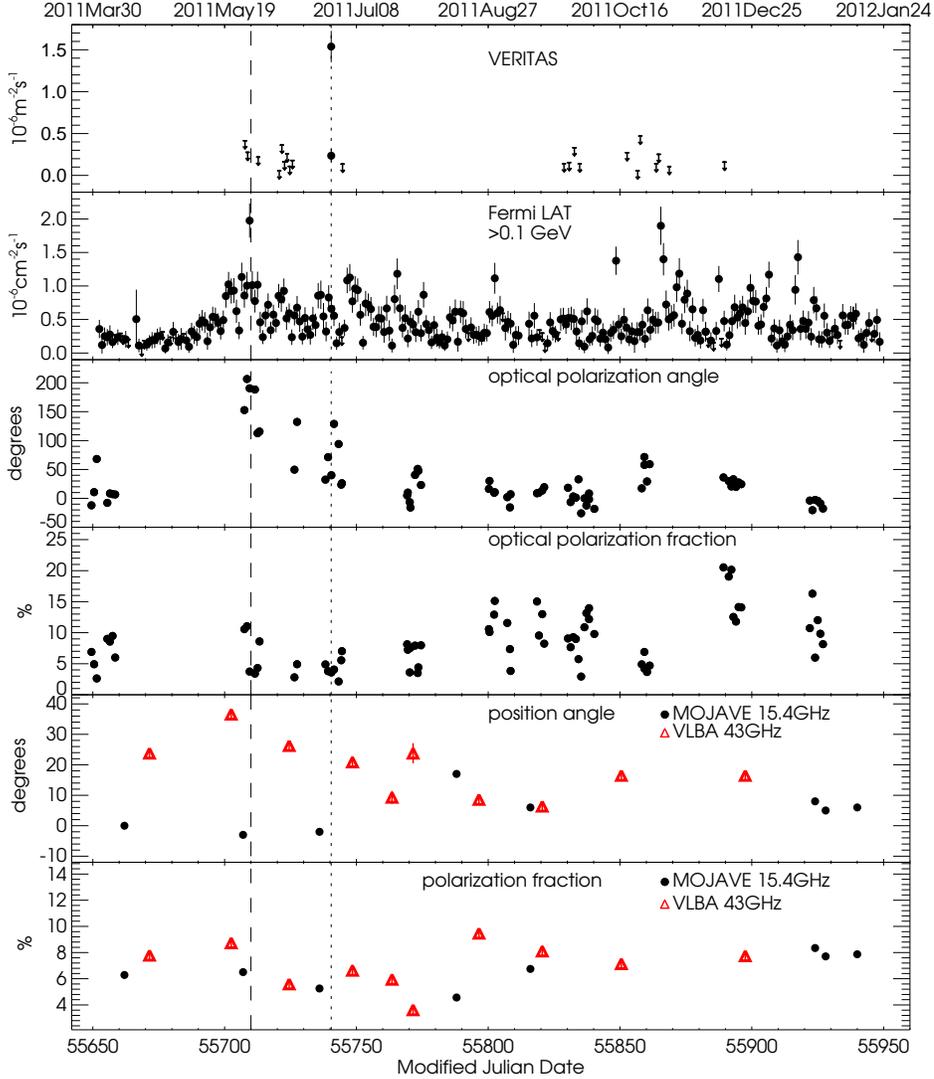}
\caption{Polarized optical and radio emission from BL Lacertae. VERITAS and {\it Fermi}-LAT light curves are also shown for comparison. The optical measurements were made at the Steward Observatory, while the radio measurements were made with the VLBA at 15.4~GHz (black dots) and 43~GHz (red triangles). The radio electric vector position angle has an uncertainty of about $\pm 3$ degrees, and the polarization fraction has an uncertainty of about 5\%. The dotted line indicates the time of the VERITAS flare, and the dashed line shows the time of the {\it Fermi}-LAT flare.}
\label{Fig plc}
\end{figure}

Fig.~\ref{Fig plc} shows results from radio polarization measurements. Although there is no significant variation in the average polarization fraction, the average polarization angle of the core appears to change before and after the TeV gamma-ray flare. However, the polarization angles for VLBA 15.4~GHz and 43~GHz do not agree with each other in earlier epochs (before the TeV gamma-ray flare). This discrepancy is likely due to the combination of the emergence of a new component, the Faraday rotation and the difference in beam size at the two frequencies. At the core, the Faraday rotation can be significant for BL Lacertae \citep{2006MNRAS.369.1596G,2007AJ....134..799J}, mostly affecting the 15.4~GHz measurements. It is also worth noting that the effects can be variable on timescales of months. 

\begin{figure}[here]
\includegraphics[width=0.48\textwidth]{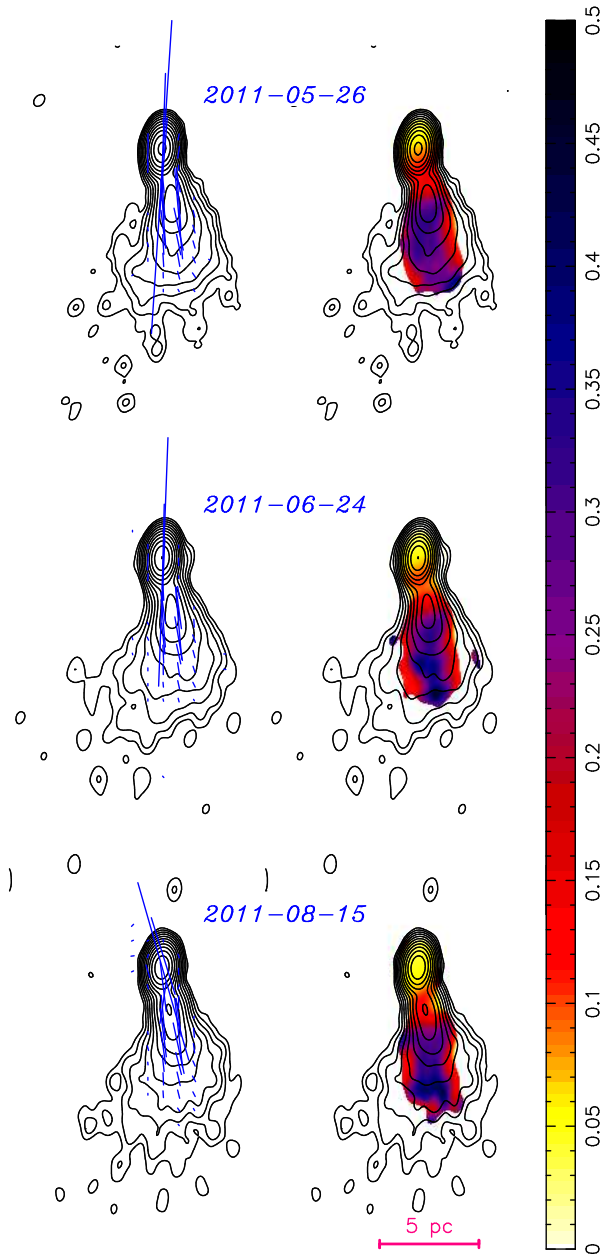}
\caption{MOJAVE 15.4 GHz VLBA images of BL Lacertae at three epochs in 2011, showing a change in core polarization after the 2011 June 28 TeV flare. The images on the left show total intensity contours, with electric polarization vectors overlaid in blue. The images on the right show total intensity contours with fractional linear polarization in color. The polarization color scale ranges from 0 to 50\%. The images have been convolved with the same Gaussian restoring beam having dimensions 0.89 mas $\times$ 0.56 mas and position angle $-8$ degrees. The base contour levels in each image are 1.3 mJy beam$^{-1}$ in total intensity and 1 mJy beam$^{-1}$ in polarization. The angular scale of the image is 1.29 pc mas$^{-1}$.}
\label{vlba15}
\end{figure}

The emergence of a new component is strongly supported by the VLBA observations at 43 GHz. Fig.~\ref{vlba43} shows a series of high-resolution images of BL Lacertae around the time of the gamma-ray flare, with the measured polarization flux and angle indicated. The results on the long-term VLBA monitoring observations will be presented elsewhere (Jorsted, Marscher et al., in prep). The new knot, K11, is discerned from the core in the images by its different polarization position angle $\chi$ ($20^\circ$ for K11 compared with $44^\circ$ for the core), even before it is clearly seen in the total intensity contours in the 2011 July 29 image. The proper motion of K11 is 0.72 mas yr$^{-1}$, which corresponds to an apparent speed of $3.6c$, with uncertainties of order 20\% owing to the short time interval over which the trajectory is followed. Although they have lower resolution, the MOJAVE images, as shown in Fig.~\ref{vlba15}, also indicate a change in the polarization of the core before and after the TeV gamma-ray flare but not in the downstream jet polarization, lending further support for the emergence of a new component associated with the gamma-ray flare.

Fig.~\ref{Fig plc} also shows results from optical polarization measurements. Again, the polarized flux does not vary significantly before and after the gamma-ray flares. However, changes in optical polarization angle are significant around the times of both GeV and TeV gamma-ray flares and between. Over the 4-day period that included the VERITAS flare, the optical polarization position angle changed by a minimum of 38.8 deg (between MJD 55738 and 55739), -31.2 deg (between MJD 55739 and 55740), and 88.8 deg (between MJD 55740 and 55741). Therefore, at a minimum, the optical polarization angle was changing by more than one degree per hour. A similar pattern is seen for the Fermi LAT flare earlier.

\section{Discussion}

For the first time, a rapid (minute-scale) TeV gamma-ray flare is seen from BL Lacertae -- this is the first such flare from an LBL. It fills an important gap between similar phenomenon observed in FSRQs and HBLs. Independent of any models, the measured decay time of the flare ($\tau_d$) requires that the size of the emitting region must be very small,
\begin{equation}
\label{eq:size}
R\le\text{c}\tau_d \delta /(1+z)\approx2.2\times 10^{13} \delta\;\text{cm},
\end{equation}
where $z$ is the redshift of the source ($z=0.069$) and $\delta$ is the Dopper factor of the jet,
\[
\delta = [\Gamma (1-\beta cos \theta)]^{-1};
\]
$\Gamma$ is the bulk Lorenz factor of the jet; and $\theta$ is the angle between the axis of the jet and the line of sight.

Another constraint for the Doppler factor is derived from the fact that the emitted gamma rays must escape $\gamma \gamma$ pair production in the emitting regions. The optical depth is given by~\citep{1995MNRAS.273..583D}
\begin{equation}
\label{eq:tau}
\tau_{\gamma \gamma}=(1+z)^{2\alpha}\delta^{-(4+2\alpha)}\frac{\sigma_T d^2_L}{5hc^2}\frac{F(\nu_0)}{T_{1/2}},
\end{equation}
where $d_L$ is the luminosity distance; $T_{1/2}$ is the doubling (not $e$-folding) time of the flare; the fiducial frequency is defined as
\[
\nu_0 = \frac{1}{\nu}\bigg ( \frac{m_ec^2}{h} \bigg )^{2} \text{,}
\]
where $\nu$ is the observed frequency of a gamma-ray photon; $F(\nu_0)$ is the observed flux at $\nu_0$. We should note that the frequency of the target photon is $\nu_0[\delta/(1+z)]$ in the jet frame and $\nu_0[\delta/(1+z)]^2$ in the observer frame. Eq.~\ref{eq:tau} is derived under the assumption that the energy spectrum of the source is approximated by introducing a power law between $\nu_0$ and $\nu_0[\delta/(1+z)]^2$, with a spectral index of $\alpha$. The fact that we detect TeV gamma-ray emission implies that the optical depth $\tau_{\gamma \gamma}$ cannot be too large. Requiring $\tau_{\gamma \gamma}<1$ leads to 
\begin{equation}
\label{eq:delta}
\delta \geq \bigg [ \frac{\sigma_T d^2_L(1+z)^{2\alpha}}{5hc^2} \frac{F(\nu_0)}{T_{1/2}} \bigg ]^{1/(4+2\alpha)} \text{.}
\end{equation}
The calculation assumes that gamma rays and target photons are both isotropic in the jet frame, so is, strictly speaking, only applicable if the gamma rays are produced via synchrotron self-Compton (SSC) scattering. For BL Lacertae, SSC may be a good approximation, as BLRs are quite weak. In this case, $d_L \approx 311 \text{Mpc}$ and $T_{1/2} \approx 9$ minutes, assuming $\Omega_m=0.27$, $\Omega_{\Lambda}=0.73$, and $H_0=70\;\text{km}\;\text{s}^{-1}\text{Mpc}^{-1}$ \citep{2011ApJS..192...16L}. At $h\nu \approx 0.9$ TeV, which is about the highest energy of all gamma rays detected within the source region, we have $\nu_0 \approx 7 \times 10^{13}$ Hz. 
Unfortunately, we did not have simultaneous IR coverage during the gamma-ray flare. Interestingly, according to \citet{2009A&A...507..769R}, the IR flux of BL Lacertae did not vary significantly (within a factor 2) during their long-term monitoring (for over 150 days) in 2007-2008 \citep[see also][]{2011ApJ...730..101A}. It is also worth noting that the synchrotron SED peak of BL Lacertae lies in the near-IR band, and the archival SEDs between near-IR and X-ray can be roughly described by a power law \citep[e.g.,][]{2003ApJ...596..847B,2010A&A...524A..43R,2011ApJ...730..101A}. The spectral index ($\alpha$) varied in the range $1.34$--$1.40$ in the frequency range $7 \times 10^{13}$--$10^{17}$ Hz. Taking $F(\nu_0)$ and $\alpha$ from the archival SEDs of BL Lacertae \citep{2009A&A...507..769R,2011ApJ...730..101A}, from Eq.~\ref{eq:delta} we found that the lower limit on $\delta$ lies in the range $13$--$17$. 

The derived lower limits on $\delta$ can be compared with other estimates. From the radio variability of BL Lacertae, \citet{2009A&A...494..527H} derived a value of $\delta = 7.3$. However, the uncertainty is expected to be large due to a number of assumptions involved in the analysis (especially in relation to the intrinsic brightness temperature). On the other hand, using a different method, \citet{2005AJ....130.1418J} arrived at a value of $\delta=7.2\pm1.1$ for different jet components, in good agreement with \citet{2009A&A...494..527H}. These values are significantly below the lower limits imposed by gamma-ray observations, perhaps implying differences between radio and gamma-ray emitting regions in the jet or a gamma-ray optical depth of $\tau_{\gamma \gamma} \gtrsim 40$. It remains to be seen whether such a strong attenuation of TeV gamma rays can be accommodated theoretically. The efforts to model the broadband SED of BL Lacertae have generally led to Doppler factors larger than 7 (e.g., \citealt{2011ApJ...730..101A}; see, however, \citealt{2004ApJ...609..576B}).

\begin{figure}[here]
\includegraphics[width=0.9\textwidth]{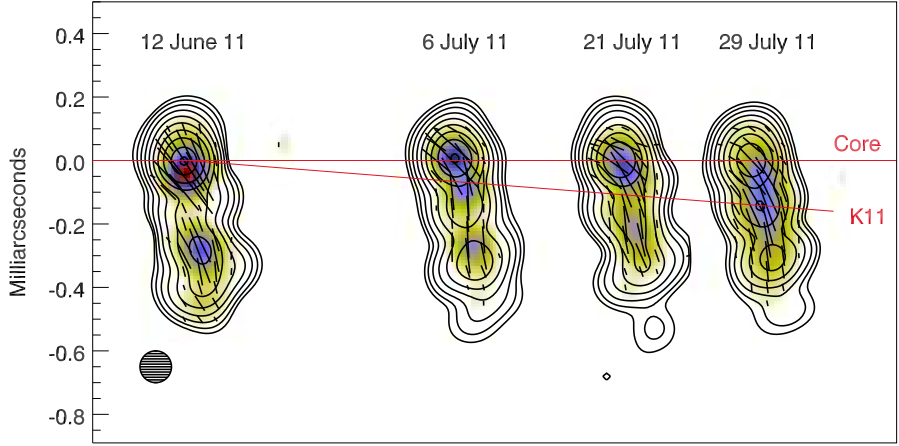}
\caption{43 GHz VLBA images of BL Lacertae at four epochs around the time of the TeV gamma-ray flare. The images are convolved with a circular Gaussian function (represented by the circle in the bottom-left corner) that has a full width at half maximum of 0.1~mas (i.e., $\sim$0.15~pc at the distance of 311~Mpc), the approximate resolution of the longest baselines of the array. Contours correspond to total intensity, with levels in factors of 2 from 0.25\%, plus an extra contour at 96\%, of the peak intensity of 2.16 Jy beam$^{-1}$. Color represents linearly polarized intensity, with maximum (black) of 0.103 Jy beam$^{-1}$ followed by red, blue, yellow, and white (no polarization detected). Red lines mark the position of the assumed stationary core and the superluminally moving knot K11, each of which has a distinct polarization position angle.}
\label{vlba43}
\end{figure}

Rapid TeV gamma-ray flaring was first observed in HBLs. It was recognized immediately that the requisite (large) Doppler factor would be problematic, because no superluminal motion had ever been seen in any of these sources \citep{2008ApJ...678...64P}. This led to the suggestion of a stratified structure of the jet that consists of a fast-moving spine and slow-moving sheath~\citep{2005A&A...432..401G}. The high-resolution polarization maps of the TeV gamma-ray HBLs have provided some evidence for such a configuration \citep{2008ApJ...678...64P}. However, for BL Lacertae, the polarization measurements do not show any stratification of the jet (for example, see Fig.~\ref{vlba43}). There is certainly no evidence for a slowly-moving sheath. Alternatively, the large Doppler factor might imply that gamma-ray production occurs in a region upstream of what is observed with the VLBA. Deceleration could explain the discrepancy between the values of the Doppler factor inferred from gamma-ray and radio observations \citep{2003ApJ...594L..27G,2007ApJ...671L..29L,2008MNRAS.383.1695S}. However, pushing the gamma-ray production region too close to the central black hole would be problematic for BL Lacertae and, even more so, for PKS 1222+21, as attenuation due to radiation from the BLRs would be strong. These difficulties might be alleviated in models that invoke subregions inside the jets that are fast moving and also sufficiently far from the black hole (\citealt{2009MNRAS.395L..29G,2012MNRAS.420..604N,2012arXiv1202.2123N}; see, however, \citealt{2011A&A...534A..86T}).
 
Of particular significance is that our multiwavelength observations of BL Lacertae might link the emergence of a superluminal knot (K11 in Fig.~\ref{vlba43}) with the TeV gamma-ray flare. The former is directly seen in the VLBA 43 GHz images, although there is a large gap in the coverage around the time of the TeV gamma-ray flare. The VLBA 15 GHz observations also show changes in the polarization angle, which supports the emergence of a new component. Based on earlier VLBA imaging, \citet{2008Natur.452..966M} argued that the core is a standing shock located well downstream (by $\sim 1$ pc) of the black hole. Their model also describes a helical magnetic field configuration upstream of the radio core, which the radiating plasma follows. This is now supported by the observed pattern of change in the optical polarization that coincides with the TeV gamma-ray flare. The new superluminal knot seems to have passed through the core on MJD 55711$\pm 15$ (2011 May 30, when the brightness centroids of the knot and core coincided), close to the time when a rapid flare was seen with the {\it Fermi}-LAT, {\it Swift} XRT and UVOT, and the Steward Observatory. 

In the model of \citet{2012arXiv1201.5402M}, the radio core is a conical shock that ends in a small shock oriented transverse to the jet axis (a Mach disk). The slow but highly compressed plasma in the Mach disk provides a highly variable local source of seed photons for inverse-Compton scattering by electrons in the faster plasma that passes across the conical shock. If a region of especially high density of relativistic electrons passes through the core, it can cause a sharp flare at gamma-ray energies and appear as a superluminal knot at radio frequencies.
Although the angular resolution of the VLBA is insufficient to measure the angular size of the knot during the observations, it is likely to have a diameter $\sim 0.07$ mas assuming that its brightness temperature is close to the value of $\sim 5\times 10^{10}$ K needed for equipartition between the energy density in relativistic electrons and that in the magnetic field \citep{1994ApJ...426...51R}. In this case, the knot interacted with the core over a period of $70\pm15$ days centered on MJD 55711 (i.e., from late April 2011 until early July 2011). Therefore, the knot would be near the end of the core region when the TeV gamma-ray flare erupted. 

\begin{figure}[here]
\includegraphics[width=0.9\textwidth]{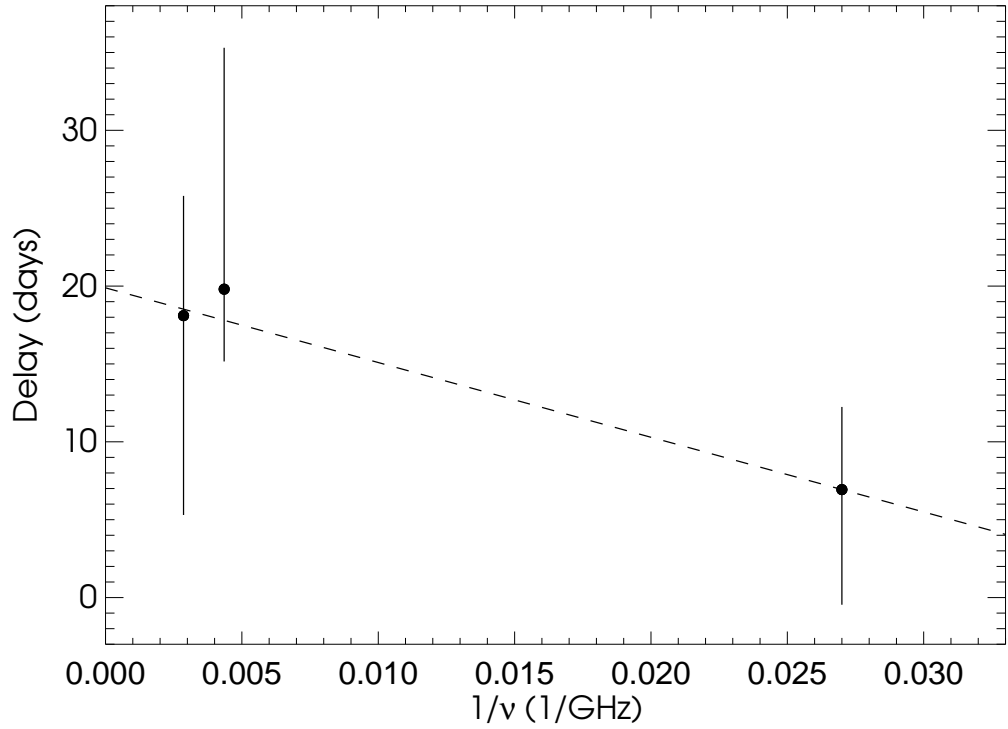}
\caption{Time delays of the radio flare from BL Lacertae. The time delays with respect to the OVRO band were determined from a likelihood code \texttt{PLIKE}. Positive delays indicate ``lead'' with respect to the OVRO band. The dashed line is drawn to guide the eye. See the text for discussion. }
\label{Fig TFreq}
\end{figure}

Alternatively, the burst of TeV gamma rays may be produced far upstream of the radio core (closer to the supermassive black hole), related to the emergence of a high-density region. As the region moves downstream, and along the helical magnetic field (as postulated by \citet{2012arXiv1201.5402M}), it produces polarized optical emission with a characteristic variation pattern of the polarization angle. When it becomes optically thin to synchrotron self-absorption, still further downstream, it is seen at successively longer wavelengths. The observed radio flare may be a manifestation of it. A delay of the radio flare by four months, with respect to the gamma-ray precursor, is in line with the fact that the radio variability of blazars generally lags the gamma-ray variability by 1--8 months \citep[e.g.,][]{2003ApJ...590...95L,2010ApJ...722L...7P,2011A&A...535A..69N,2012arXiv1204.3589L}. Theoretically, the optical-depth effect should lead to a $\nu^{-1}$ dependence of the time lag, as the core remains optically thick to synchrotron self-absorption up to a distance (from the black hole) $r_{c}\propto\nu^{-1}$ \citep{1979ApJ...232...34B}. Unfortunately, the measurements (as shown in Fig.~\ref{Fig TFreq}) are not of sufficient quality to confirm such a frequency dependence.

The lack of similarly rapid change of significant amplitude at other wavelengths is likely due to inadequate sampling. In other words, the TeV gamma-ray flare is so rapid that pointed instruments were unlikely to be observing the source at the right time, while for other instruments (e.g., the {\it Fermi} LAT) it is difficult to accumulate adequate statistics. Nevertheless, around the time of the TeV gamma-ray flare, there is evidence for flux variations at optical and UV wavelengths, which would represent a response of the synchrotron emission to the VHE gamma-ray flaring. 

\acknowledgments
QF and WC wish to thank Dimitrios Giannios for useful comments on the manuscript. The work of the VERITAS Collaboration was supported by grants from the U.S. Department of Energy Office of Science, the U.S. National Science Foundation and the Smithsonian Institution, by NSERC in Canada, by Science Foundation Ireland (SFI 10/RFP/AST2748) and by STFC in the U.K. We acknowledge the excellent work of the technical support staff at the Fred Lawrence Whipple Observatory and at the collaborating institutions in the construction and operation of the instrument. 

The Boston University effort was supported in part by NASA through {\it Fermi} grants NNX08AV65G, NNX08AV61G, NNX09AT99G, and NNX11AQ03G, and by NSF grant AST-0907893.

The OVRO 40-m monitoring program is supported in part by NASA grants NNX08AW31G and NNX11A043G, and NSF grants AST-0808050 and AST-1109911.

The MOJAVE project is supported under NASA-Fermi grant NNX08AV67G. The National Radio Astronomy Observatory is a facility of the National Science Foundation operated under cooperative agreement by Associated Universities, Inc. YYK and KVS were partly supported by the Russian Foundation for Basic Research (project 11-02-00368, 12-02-33101) and the basic research program ``Active processes in galactic and extragalactic objects'' of the Physical Sciences Division of the Russian Academy of Sciences. YYK was supported by the Dynasty Foundation.

The Submillimeter Array is a joint project between the Smithsonian Astrophysical Observatory and the Academia Sinica Institute of Astronomy and Astrophysics and is funded by the Smithsonian Institution and the Academia Sinica.

The Steward Observatory spectropolarimetric monitoring project is supported by {\it Fermi} Guest Investigator grants NNX08AW56G and NNX09AU10G. 

The Mets\"ahovi team acknowledges the support from the Academy of Finland to observing projects (numbers 212656, 210338, 121148, and others).

This research has made use of data from the MOJAVE database that is maintained by the MOJAVE team \citep{2009AJ....137.3718L}, the Swinburne University of Technology software correlator that is developed as part of the Australian Major National Research Facilities Programme and operated under license, and the LAT public archive that is maintained by the Fermi Science Support Center.

\clearpage

\clearpage

\clearpage

\clearpage

\clearpage

\clearpage

\clearpage

\clearpage

\clearpage

\clearpage

\clearpage

\end{document}